\documentclass[final,1p,times, compress, sort]{elsarticle}
\usepackage{amssymb}
\usepackage{siunitx}
\journal{NIMA}
\usepackage{hyperref}
\hypersetup{
    colorlinks=true,
    linkcolor=blue,
    filecolor=magenta,      
    urlcolor=cyan,
}
\usepackage[switch]{lineno}

\begin{document}

\begin{frontmatter}



\title{Defects and acceptor removal in $^{60}$Co $\gamma$-irradiated \textit{p}-type silicon}


\author[1]{Anja Himmerlich\corref{cor1}}
\ead{anja.himmerlich@gmx.de}
\author[4]{N\'{u}ria Castell\'{o}-Mor}
\author[1]{Esteban Curr\'{a}s-Rivera}
\author[{1}]{Yana Gurimskaya}
\author[{1}]{Isidre Mateu}
\author[1]{Michael Moll}
\author[1]{Karol Pawel Peters}
\author[1,5]{Niels Sorgenfrei}
\author[1]{Moritz Wiehe}
\author[2]{Andrei Nitescu}
\author[2]{Ioana Pintilie}
\author[3]{Eckhart Fretwurst}
\author[3]{Chuan Liao}
\author[3]{Jörn Schwandt}

\cortext[cor1]{Corresponding author}
\affiliation[1]{organization={European Organization for Nuclear Research, CERN},%Department and Organization
            addressline={Esplanade des Particules 1},
            city={Geneva},
            postcode={1211}, 
            country={Switzerland}}

\affiliation[2]{organization={National Institute of Materials Physics, NIMP},%Department and Organization
            addressline={Str. Atomistilor 105 bis}, 
            city={Bucharest},
            postcode={RO-77125}, 
            country={Romania}}

\affiliation[3]{organization={Institute for Experimental Physics, University of Hamburg},%Department and Organization
            addressline={Luruper Chaussee 149}, 
            city={Hamburg},
            postcode={22761}, 
            country={Germany}}

\affiliation[4]{organization={Instituto de Física de Cantabria (IFCA), CSIC - Universidad de Cantabria},%Department and Organization
            addressline={Avenida de los Castros, s/n}, 
            city={Santander},
            postcode={39005}, 
            country={Spain}}

\affiliation[5]{organization={Institute of Physics, Albert-Ludwigs-Universitaet Freiburg},%Department and Organization
            addressline={Hermann-Herder-Strasse 3}, 
            city={Freiburg im Breisgau},
            postcode={79104}, 
            country={Germany}}

\begin{abstract}
Boron-doped silicon detectors used in high radiation environments like the future HL-LHC show a degradation in device performance due to the radiation induced deactivation of the active boron dopant. This effect, known as the so-called Acceptor Removal Effect (ARE), depends on particle type, particle energy and radiation dose and is usually explained by the formation of boron-interstitial - oxygen-interstitial (B{$_\text{i}$}O{$_\text{i}$}) defects that induce a donor-type defect level in the upper part of the Si band gap. Here we present defect characterization studies using Thermally Stimulated Current technique (TSC) and Deep Level Transient Spectroscopy (DLTS) on a set of epitaxially grown \textit{p}-type silicon diodes of different resistivity, irradiated with $^{60}$Co $\gamma$-rays. We used the defect parameters (activation energy, charge carrier capture cross sections and defect concentration) obtained from DLTS experiments for modeling the corresponding TSC spectra, and subsequently compared those with the experimental TSC results. This approach shows that the di-vacancy which is well characterized by DLTS correlates with the so-far unspecified charge emission signal of the X-defect that partially overlaps with the B{$_\text{i}$}O{$_\text{i}$} peak in TSC spectra.
Additionally, in order to evaluate the impact of B{$_\text{i}$}O{$_\text{i}$} defect formation on the macroscopic properties of the device, we compared the B{$_\text{i}$}O{$_\text{i}$} defect concentration with the change in the effective carrier concentration \textit{N}$_\text{eff}$ obtained from \textit{C-V} measurements. It shows that the variations in \textit{N}$_\text{eff}$ are about twice the changes in the B{$_\text{i}$}O{$_\text{i}$} concentration, which is in perfect consistency with the assumption of boron deactivation by the formation of the B{$_\text{i}$}O{$_\text{i}$} donor in irradiated \textit{p}-type Si. 

\end{abstract}




\end{frontmatter}

%\linenumbers

\section{Introduction}
Boron-doped silicon detectors (e.g. $n^{+}-p$ diodes, Low Gain Avalanche Detectors (LGADs) or HV-CMOS devices) used in high radiation environments like the future HL-LHC at CERN are impacted in their performance due to the radiation induced deactivation of the active boron dopant. This effect is known as the so-called Acceptor Removal Effect (ARE) and depends on particle type, particle energy and radiation dose \cite{Wunstorf1996,Terada1996, Kramberger2015, Sadrozinski2018, Kramberger2018, Ferrero2019NIMA, Moll2018IEEE, Moll-Vertex2019}. Briefly summarized, one assumes that the impinging high-energy particles induce a displacement damage in the silicon lattice creating Frenkel pairs composed of a silicon atom on an interstitial site (I) as well as a vacancy (V). Vacancies usually show low mobility at low temperature and mainly form vacancy-oxygen complexes or multi-vacancy defects (V$_\text{2}$, V$_\text{3}$, ...). Si-interstitials however are very mobile even at cryogenic temperatures and interact via the Watkins replacement mechanism with impurities and dopants \cite{Watkins2000MatSciSem}. In boron-doped Si the two main interaction pathways result in the formation of boron-interstitials (B$_\text{i}$) and carbon-interstitials (C$_\text{i}$) which further interact with oxygen to boron-interstitial - oxygen-interstitial (B{$_\text{i}$}O{$_\text{i}$}) or carbon-interstitial - oxygen-interstitial (C{$_\text{i}$}O{$_\text{i}$}) defects \cite{Kimerling1989, Mooney1977, Moll-Vertex2019}. The C{$_\text{i}$}O{$_\text{i}$} defect is known to be neutral in the space charge region, while the B{$_\text{i}$}O{$_\text{i}$} forms a donor-type defect level in the upper part of the Si band gap that introduces positive space charge \cite{Kimerling1989}. Consequently, the formation of the B{$_\text{i}$}O{$_\text{i}$} defect deactivates in total two active boron dopants - one due to the inclusion of the B{$_\text{i}$}O{$_\text{i}$} defect and one by counterbalancing the negative space charge of a second boron by the positive space charge of the B{$_\text{i}$}O{$_\text{i}$}. Therefore, B{$_\text{i}$}O{$_\text{i}$} formation should correlate with a factor of two with the change in the effective space charge concentration \textit{N}$_\text{eff}$. Here it should be mentioned that in recent publications also immobile boron-substitutionals that are capturing Si-interstitials forming a so-called B$_\text{Si}$Si$_\text{i}$ defects are considered to explain the ARE \cite{Lauer2020PSSA}. However, in this publication we will follow the so far widely accepted assumption of a B{$_\text{i}$}O{$_\text{i}$} defect structure. \\
Besides the discussed point-like defects, high energetic radiation induced displacements also create cluster-like defects that significantly impact the performance of silicon detectors \cite{Moll2018IEEE, Pintilie2009NIMA, Pintilie2008APL}. However, by performing radiation studies using low-energy electrons or $\gamma$-radiation the creation of large cluster complexes is very unlikely, and thus, it becomes possible to directly correlate changes in the macroscopic device properties, like the change in \textit{N}$_\text{eff}$, with changes in concentrations of point-like defects like the B{$_\text{i}$}O{$_\text{i}$} \cite{Watkins2000MatSciSem, Pintilie2009NIMA, Chuan2023gamma}. \\
In this work, we present defect spectroscopy studies in combination with electrical characterization as well as theoretical modeling approaches on $\gamma$-irradiated \textit{p}-type silicon diodes of different resistivity with the intention to further deepen and intensify the knowledge of radiation induced point-like defects and their impact on the performance of Si based devices. 
%%%%%%%%%%%%%%%%%%%%%%%%%%%%%%%%%%%%%%%%%%
\begin{table}[b!] 
\centering
\begin{tabular}{l c c  c c   }
\hline
 & resitivity     & nominal dose      & \multicolumn{2}{c}{IR }      \\
 & [$\Omega$cm]   & [MGy]          &  \multicolumn{2}{c}{ [10$^{6}$ cm$^{-3}$/Gy]}  \\
 &     & & B{$_\text{i}$}O{$_\text{i}$} & C{$_\text{i}$}O{$_\text{i}$}  \\
\hline
EPI-06-DS-67 & 50 & 0.1  & 6.3  & 1.1 \\
EPI-06-DS-69 & 50 & 0.2  & 6.7 & 1.4  \\
EPI-06-DS-82 & 50 & 1  & 5.8   & 1.5 \\
EPI-06-DS-84 & 50 & 2  & 5.5  & 1.4  \\
\hline
EPI-10-DS-78 & 250 & 0.1  & 4.4 & 4.4  \\
EPI-10-DS-80 & 250 & 0.2  & 4.5  & 4.7  \\
EPI-10-DS-82 & 250 & 1  &  4.8  & 5.2  \\
EPI-10-DS-94 & 250 & 2  &  4.7  & 5.1  \\
\hline
\end{tabular}
\caption{Sample overview. Given are the sample numbers, resistivity of the samples and dose rates as well as introduction rates (IR) for the B{$_\text{i}$}O{$_\text{i}$} and C{$_\text{i}$}O{$_\text{i}$} defects. The latter were calculated by using defect concentrations obtained by DLTS measurements.}
\label{tab:samples}
\end{table}
%%%%%%%%%%%%%%%%%%%%%%%%%%%%%%%%%%%%%%%%%%
\section{Materials and methods}
\label{sec:MatMet}
The experimental studies were performed on a set of boron-doped silicon pad diodes with $n^{+}-p-p^{+}$ structure manufactured by CiS (Forschungsinstitut für Mikrosensorik GmbH, Erfurt, Germany) \cite{CIS}. They consist of an 50\,\unit{\micro\meter} boron-doped bulk layer of either 50\,$\Omega$cm or 250\,$\Omega$cm resistivity epitaxially grown on a low resistivity substrate. The active area of the devices is \num{6.927E-2}\,cm$^{2}$. More details about the diodes can be found in Ref. \cite{Gurimskaya2020} and \cite{Himmerlich2022}. The devices were $\gamma$ irradiated at the $^{60}$Co source of the Ruder Boskovic Institute (RBI) in Zagreb, Croatia with doses in the range of 0.1 to 2\,MGy \cite{Chuan2023gamma}. An overview of the sensors and radiation doses is given in Tab.\ref{tab:samples}. After irradiation, performed at room temperature, the sensors were not intentionally annealed. They were kept at temperatures below -\,20$^\circ$C also during the transport. \\
In order to investigate radiation induced macroscopic changes of the device properties, Capacitance-Voltage (\textit{C-V}, measurement frequency 10\,kHz, parallel mode) and Current-Voltage (\textit{I-V}) measurements were performed at -\,20$^\circ$C. During all measurements the guard ring of the diodes was connected to ground.\\
\textit{C-V} measurements were used to extract the effective carrier concentrations \textit{N}$_\text{eff}$ of the irradiated diodes using the following equations \cite{Schroder-book}:
\begin{equation}
    N_\text{eff} = \frac{2}{\epsilon_r\,\epsilon_0\,q_0\,A^2\,d(1/C^2)/dV}
    \label{eq:slope}
\end{equation}
\begin{equation}
    N_\text{eff} = \frac{2\epsilon_r\,\epsilon_0}{q_0\,w^2}\,V_\text{depl}
    \label{eq:kink}
\end{equation}
with: $\epsilon$ - the dielectric constant, \textit{q}$_\text{0}$ - the elementary charge, \textit{A} - the electrode area of the diode, \textit{w} - the thickness of the depleted region and \textit{d(1/C$^\text{2}$)/dV} - the slope of the 1/\textit{C$^\text{2}$} versus voltage curve.

To characterize the radiation induced defects in the investigated diodes, Deep-Level-Transient Spectroscopy (DLTS) and Thermally Stimulated Current technique (TSC) were used.  Both methods can be used to extract defect parameters like thermal activation energy \textit{ E}$_\text{a}$ of defect levels, capture cross sections for electrons and holes $\sigma_\text{n,p}$ and defect concentration \textit{N}$_\text{T}$. While DLTS technique is based on analyzing  the measured capacitance transients following injection pulses for filling the traps at each temperature step, in TSC the quasi-equilibrium currents generated by emission of charge from the  traps filled at low temperature (single shot injection) are used to determine the properties and concentration of defects. In TSC, the occupation of the defect levels was done by cooling down the diodes under reverse bias (UR) to a filling temperature \textit{T}$_\text{fill}$ where the filling of the traps was performed either with majority carriers (holes) by setting the bias voltage to around 0\,V for a certain filling time \textit{t}$_\text{fill}$, or with both, minority and majority carriers (electrons and holes) by applying a forward bias of +\,20\,V, corresponding to a filling current of about 1\,mA which was set as compliance for the source meter. A more detailed description of both techniques can be found in Ref.\,\cite{Weiss1988, Pintilie2001, Himmerlich2022, Liao2022IEEE, Chuan2023electron}.\\
In order to compare defect levels identified by DLTS with those measured with TSC we used the defect parameters obtained by DLTS to model TSC spectra by using a Python-based analysis software \textit{pytsc}. Within \textit{pytsc} the diode is treated as one-dimensional device with a homogeneous dopant and defect distribution inside the sensor volume. The discharging TSC current is given by the following equation \cite{Moll-thesis}:
\begin{equation}
    I_{TSC}(t) = q_{0}\,A\,\sum^{n}_{i=1}\,\Biggl\lbrack\,\sum_{defects} \frac{e_n(t)\,n_\text{T}(t) + e_p(t)\,p_\text{T}(t)}{2}\Biggr\rbrack \Delta z_{i}
    \label{eq:ITSC}
\end{equation}
where \textit{e$_\text{n}$(t)} and \textit{e$_\text{p}$(t)} are the emission rates for electrons and holes while \textit{n}$_\text{T}$\textit{(t)} and \textit{p}$_\text{T}$\textit{(t)} are the fraction of defect states occupied by electrons or holes. For the simulation the sensor bulk volume is sliced into N differential parts of thickness $\Delta$z and n is the number of fully depleted slices. This number depends on the applied bias voltage as well as the effective carrier concentration \textit{N}$_\text{eff}$, both defining the depletion depth (equivalent to equation \ref{eq:kink}). \textit{N}$_\text{eff}$ can set constant or dynamic (\textit{N}$_\text{eff}$\,=\,\textit{N}$_\text{eff,0}$\,+\,\textit{N}$_\text{T}$(t)) within \textit{pytsc}. For the results presented in this paper \textit{N}$_\text{eff}$ was always set constant. \\
The emission rate for electrons or holes from a single trap is given by:
\begin{equation}
    e_{n,p}(t) = \Gamma_{PF}(E,T)\cdot \xi \cdot \sigma_{n,p} \cdot exp(-\frac{E_a}{k_B T})
\end{equation}
with $\Gamma_{PF}$\textit{(E,T)} giving the enhanced emission probability, in dependence of the position dependent electric field distribution \textit{E} and the temperature \textit{T}, due to the Poole-Frenkel effect, and
$\mathrm{\xi=B\cdot\frac{m^{*}_{dC,dV}}{m_0}\cdot T^2\cdot \nu^{th}_{n,p}}$, including a constant B, the effective mass m$^\text{*}_\text{n,p}$ as well as thermal velocity $\nu^\text{th}_\text{n,p}$ of the carriers. $\sigma_{n,p}$ is the cross section and \textit{E}$_\text{a}$ is the activation energy. In the modeling given in this publication the influence of the Poole-Frenkel effect is not included, so $\Gamma_{PF}$\textit{(E,T)} was set to one. \\
The number of defect states occupied by electrons \textit{n}$_\text{T}$\textit{(t)} (and equivalent for holes \textit{p}$_\text{T}$\textit{(t)}) at a certain time t is given by:
\begin{equation}
    n_\text{T}(t) = N_\text{T}(t_0) \cdot exp\Bigl\lgroup\int_{t_0}^{t}e_n (t') dt'\Bigr\rgroup
\end{equation}
For the modeling with \textit{pytsc} presented in this publication, as absolute trap concentrations \textit{N}$_\text{T}$, capture cross sections and activation energies, experimentally determined values from corresponding DLTS measurements were used.

%, w(t) - the depletion depth and $\beta$(t) = dT/dt - the constant heating rate for the temperature scan.\\
%\begin{equation}
%    I_{TSC}(t) = q_{0}\,A\,\int^{w(t)}_0 \sum_{defects} \frac{e_n(t)\,n_t(t) + e_p(t)\,p_t(t)}{2} dx
%    \label{eq:ITSC}
%\end{equation}

\section{Results and Discussion}
\subsection{Electrical Characterization}
\begin{figure}[t!]
    \centering
    \includegraphics[width=1\columnwidth]{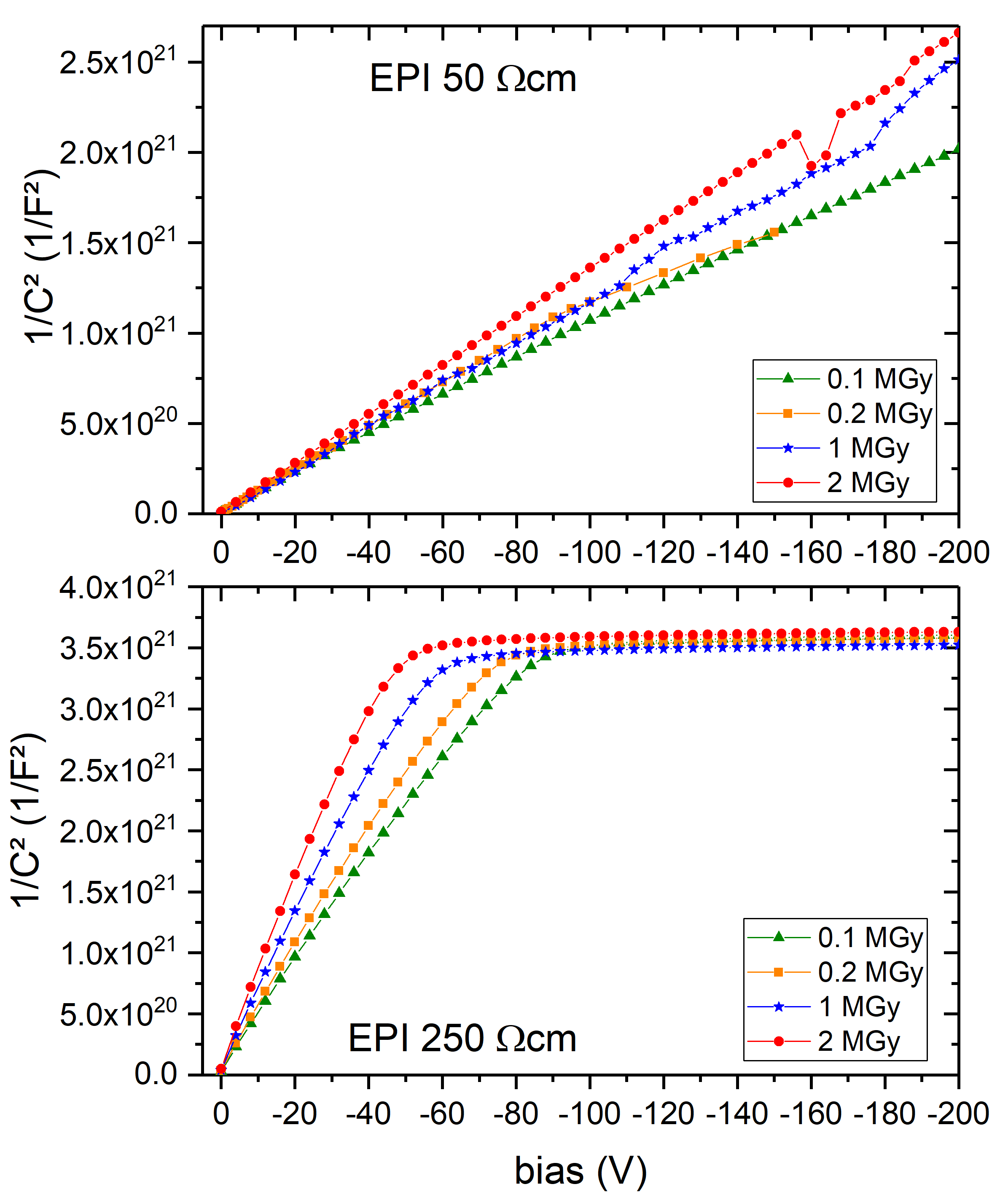}
    \caption{\textit{C-V} measurements of $^{60}$Co gamma irradiated EPI diodes of different resistivity (top: 50\,$\Omega$cm, bottom: 250\,$\Omega$cm). The diodes were measured at \text{-\,20}$^{o}$\,C and with a frequency of 1\,kHz.} 
    \label{fig:CV}
\end{figure}

All samples were electrically characterized before and after irradiation. In Fig.\ref{fig:CV} capacitance-voltage (\textit{C-V}) measurements performed on 50\,$\Omega$cm (top) and 250\,$\Omega$cm (bottom) EPI diodes are plotted. The curves were recorded after $^{60}$Co gamma irradiation with different doses. Due to the high doping, the 50\,$\Omega$cm sensors could not be fully depleted in the applied voltage range. The full depletion voltage \textit{V}$_\text{depl}$ of the 250\,$\Omega$cm sensors, extracted from the kink in the curves, decreases from about -\,82\,V after 0.1\,MGy irradiation to about -\,45\,V after 2\,MGy irradiation.   
From the \textit{C-V} measurements the effective carrier concentrations \textit{N}$_\text{eff}$ were extracted using equation \ref{eq:slope} and \ref{eq:kink}.
The values are plotted in Fig.\,\ref{fig:concentr-Neff} and will be discussed in section\,\ref{subsec: DLTS}. \\
Current-voltage (\textit{I-V}) measurements on the $^{60}$Co gamma irradiated EPI diodes (not illustrated here) have shown an increase of the leakage current with increasing radiation dose.
%--------------------------------
\begin{figure}[b!]
    \centering
    \includegraphics[width=1\columnwidth]{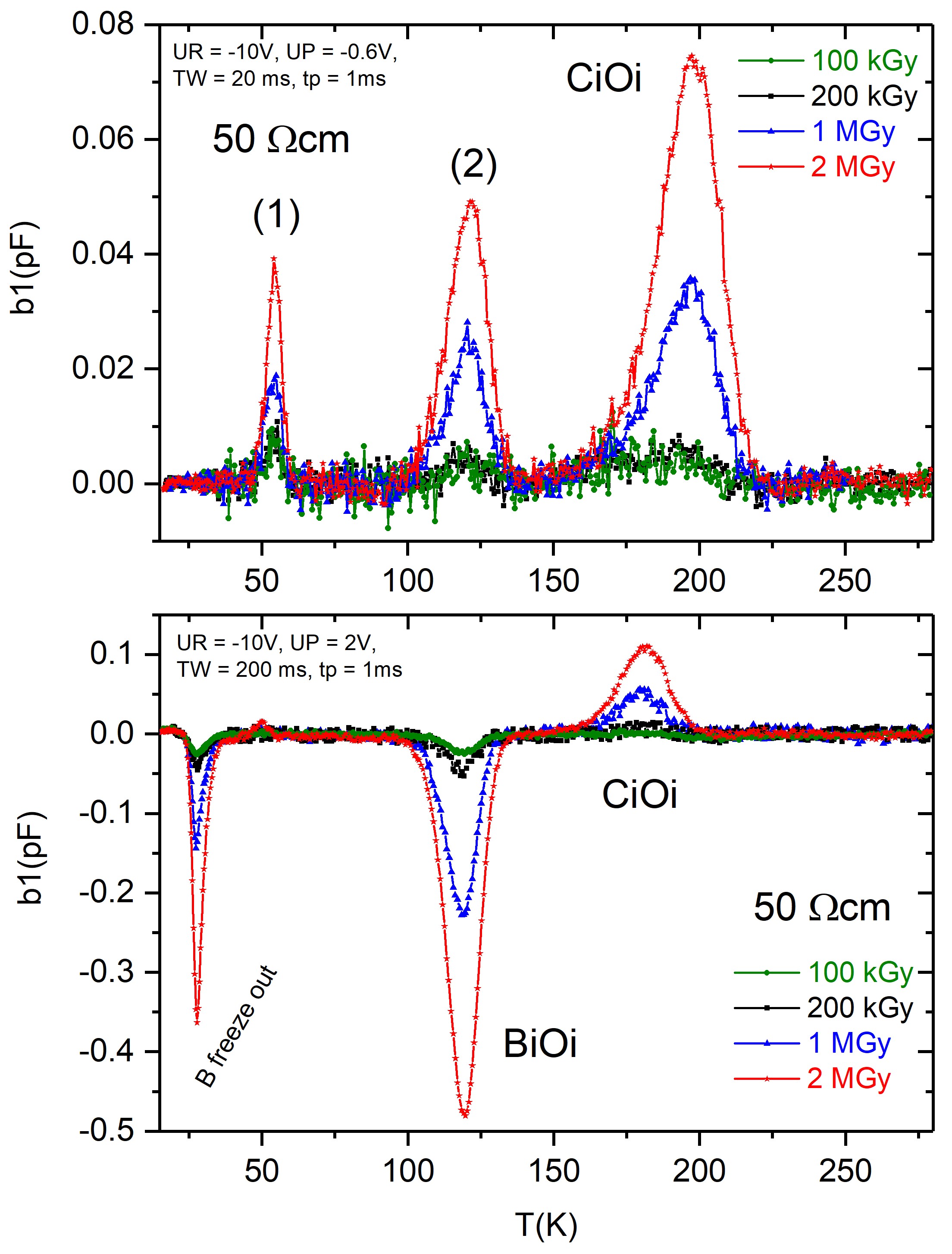}
    \caption{DLTS measurements on 50\,$\Omega$cm $^{60}$Co $\gamma$-irradiated EPI diodes. top: majority carrier injection, bottom: majority and minority carrier injection} 
    \label{fig:50Ohm-DLTS}
\end{figure}
%--------------------------------
\begin{figure}[ht!]
    \centering
    \includegraphics[width=1\columnwidth]{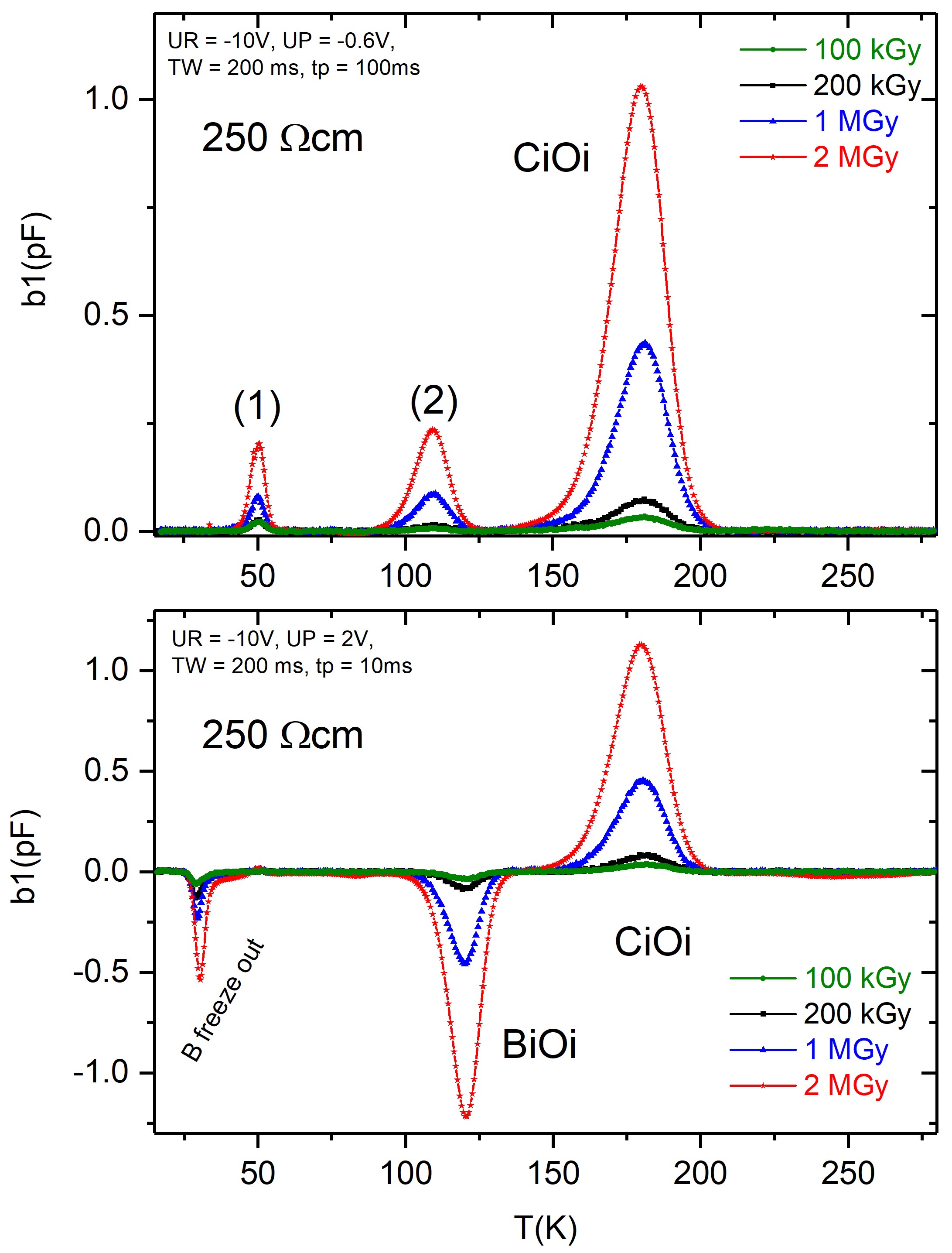}
    \caption{DLTS measurements on 250\,$\Omega$cm $^{60}$Co $\gamma$-irradiated EPI diodes. top: majority carrier injection, bottom: majority and minority carrier injection} 
    \label{fig:250Ohm-DLTS}
\end{figure}
%--------------------------------
\subsection{DLTS studies}
\label{subsec: DLTS}
In Fig. \ref{fig:50Ohm-DLTS} and \ref{fig:250Ohm-DLTS} the measured DLTS spectra of the $\gamma$-irradiated 50\,$\Omega$cm and 250\,$\Omega$cm sensors are presented. The upper spectra in Fig. \ref{fig:50Ohm-DLTS} and Fig. \ref{fig:250Ohm-DLTS}  were obtained by applying a pulse voltage UP of -0.6\,V corresponding to majority carrier injection (only hole traps are detected), while the lower spectra were obtained after applying UP\,=\,+\,2\,V leading to majority and minority carrier injection (hole and electron traps can be detected).

For both sensor types four pronounced defect levels are detected, three hole traps and one electron trap, labeled as (1), (2), C$_\text{i}$O$_\text{i}$ (carbon-intersital - oxygen-interstitial) and B$_\text{i}$O$_\text{i}$ (boron-interstitial - oxygen-interstitial), respectively. By analysing the Arrhenius-plots resulted from the DLTS spectra corresponding to different time windows (TW) the following defect parameters were obtained for the detected levels:

\begin{center}
peak (1): E$_\text{v}$\,+\,0.09\,eV and $\sigma_\text{p}$: 
2$\cdot$10$^{-14}$\,cm$^2$\\
peak (2): E$_\text{v}$\,+\,0.19\,eV and $\sigma_\text{p}$: 4$\cdot$10$^{-16}$\,cm$^2$\\
C$_\text{i}$O$_\text{i}$: E$_\text{v}$\,+\,0.36\,eV and $\sigma_\text{p}$:2$\cdot$10$^{-15}$\,cm$^2$\\
B$_\text{i}$O$_\text{i}$: E$_\text{c}$\,-\,0.25\,eV and $\sigma_\text{n}$: 6$\cdot$10$^{-15}$\,cm$^2$\\
\end{center}
In Addition, for peak (2) direct capture cross section measurements were performed, giving a value of $\sigma_\text{p}$\,=\,7$\cdot$10$^{-16}$\,cm$^2$. 

%--------------------------------
\begin{figure}[b!]
    \centering
    \includegraphics[width=1\columnwidth]{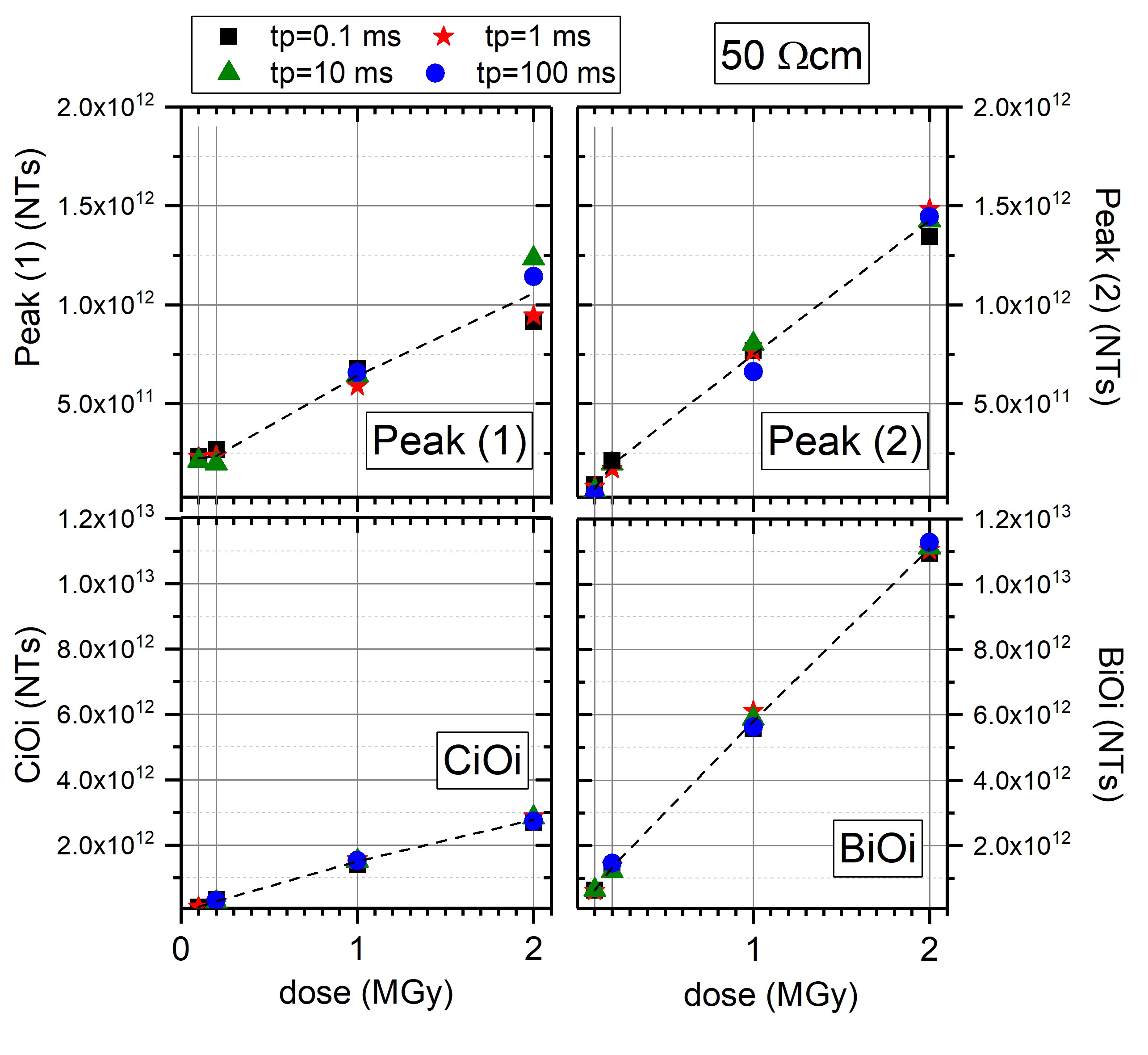}
    \caption{Concentrations of the four dominant defects measured by DLTS on 50\,$\Omega$cm $^{60}$Co gamma irradiated EPI diodes. Each diode was measured by varying the pulse duration \textit{t}$_\text{p}$ between 0.1\,ms and 100\,ms. The dotted lines indicate the mean concentration values of all $t_\text{p}$ at each radiation dose and guides the eye between measured doses. } 
    \label{fig:50Ohm-concentr}
\end{figure}
%--------------------------------
For the C$_\text{i}$O$_\text{i}$ and B$_\text{i}$O$_\text{i}$ defect the values are in good agreement with data from the literature. Also for peak (1) and peak (2) an assignment can be given by comparing the defect parameters to DLTS data in the literature \cite{Watkins2000MatSciSem, Mooney1977, Seebauer-book, Zangenberg2002, Zangenberg2005AP, Aha2021PSSa, Markevich2005JCM, Markevich2016SSP, Gusakov2017PSSa}. Thereby, peak (1) may be related to an di-self-interstial - oxygen-interstitial (I$_{2}$O) defect, while the characteristics of peak (2) are comparable to a vacancy related defect level: the single-positive charge state of the di-vacancy V$_\text{2}$(0/+) . \\
%--------------------------------
\begin{figure}[b!]
    \centering
    \includegraphics[width=1\columnwidth]{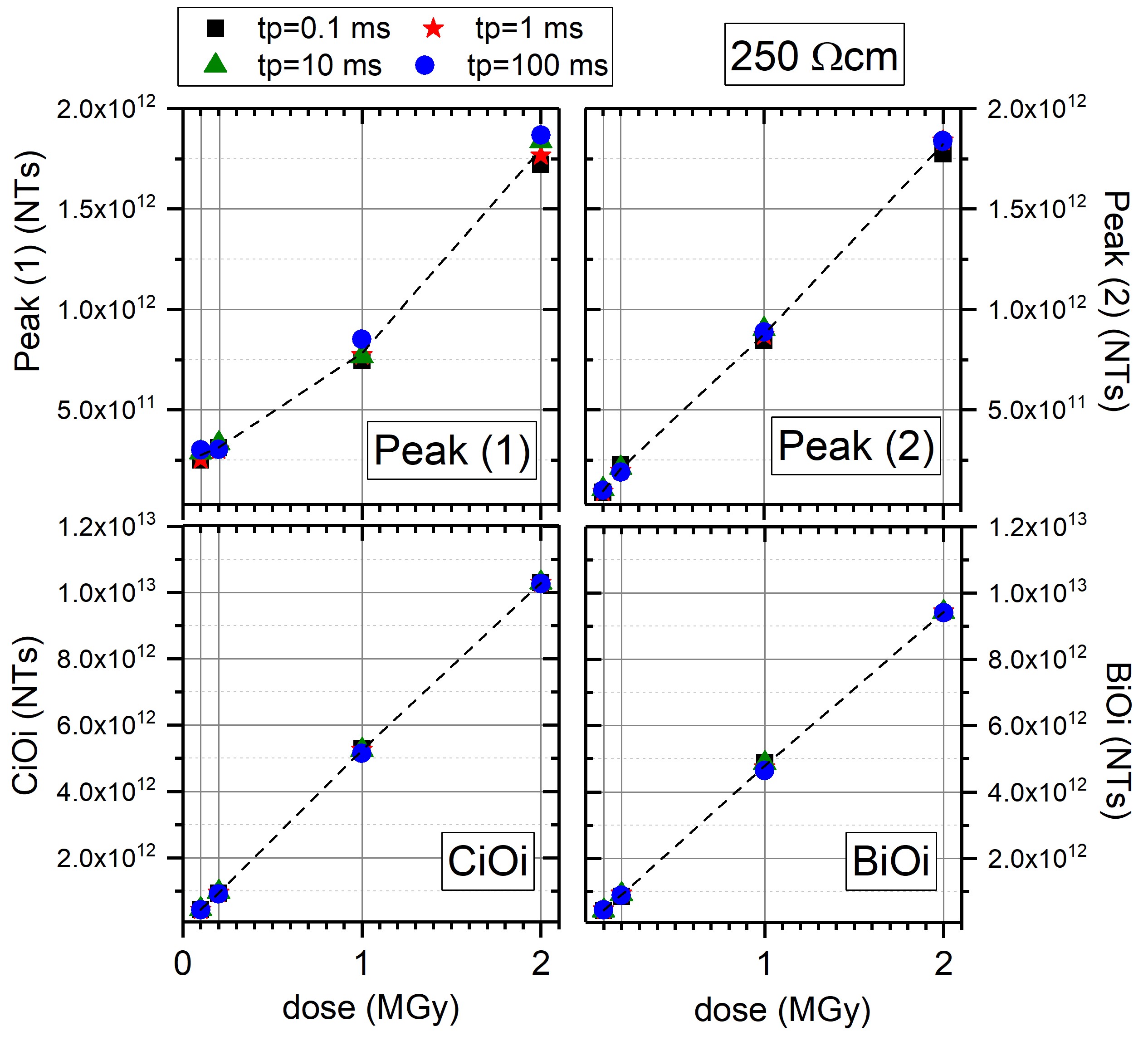}
    \caption{Concentrations of the four dominant defects measured by DLTS on 250\,$\Omega$cm $^{60}$Co gamma irradiated EPI diodes. Each diode was measured by varying the pulse duration \textit{t}$_\text{p}$ between 0.1\,ms and 100\,ms. The dotted lines indicate the mean concentration values of all $t_\text{p}$ at each radiation dose and guides the eye between measured doses. } 
    \label{fig:250Ohm-concentr}
\end{figure}
%--------------------------------
The extracted concentrations for the detected defects as function of radiation dose are given in Fig.\ref{fig:50Ohm-concentr} and Fig.\,\ref{fig:250Ohm-concentr} for the 50\,$\Omega$cm and the 250\,$\Omega$cm sensors, respectively. The different symbols for one radiation dose correspond to measurements with different duration of the injection pulse (t$_\text{p}$) from 0.1\,ms up to 100\,ms. 
No significant dependence of the extracted defect concentrations on the pulse time was found. The defect concentrations of peak (2), C$_\text{i}$O$_\text{i}$ and B$_\text{i}$O$_\text{i}$ increase linearly with radiation dose, while for the high resistivity diode, the increase of the defect concentration of peak (1) is steeper at a higher radiation dose.  In table \ref{tab:samples} the corresponding introduction rates for the B$_\text{i}$O$_\text{i}$ and C$_\text{i}$O$_\text{i}$ are listed.\\
Concerning differences in the absolute defect concentrations between the 50\,$\Omega$cm and the 250\,$\Omega$cm diodes, it can be observed, that for peak (1) and peak (2) the defect concentrations are slightly higher for the 250\,$\Omega$cm diodes. In the 50\,$\Omega$cm diodes the B$_\text{i}$O$_\text{i}$ concentration dominates due to the high boron content in the samples. 
The C$_\text{i}$O$_\text{i}$ concentration is rather low in these diodes, especially compared to those of 250\,$\Omega$cm where the C$_\text{i}$O$_\text{i}$ to B$_\text{i}$O$_\text{i}$ ratio is $\geq$\,1.     

\begin{figure}[htb]
    \centering
    \includegraphics[width=1\columnwidth]{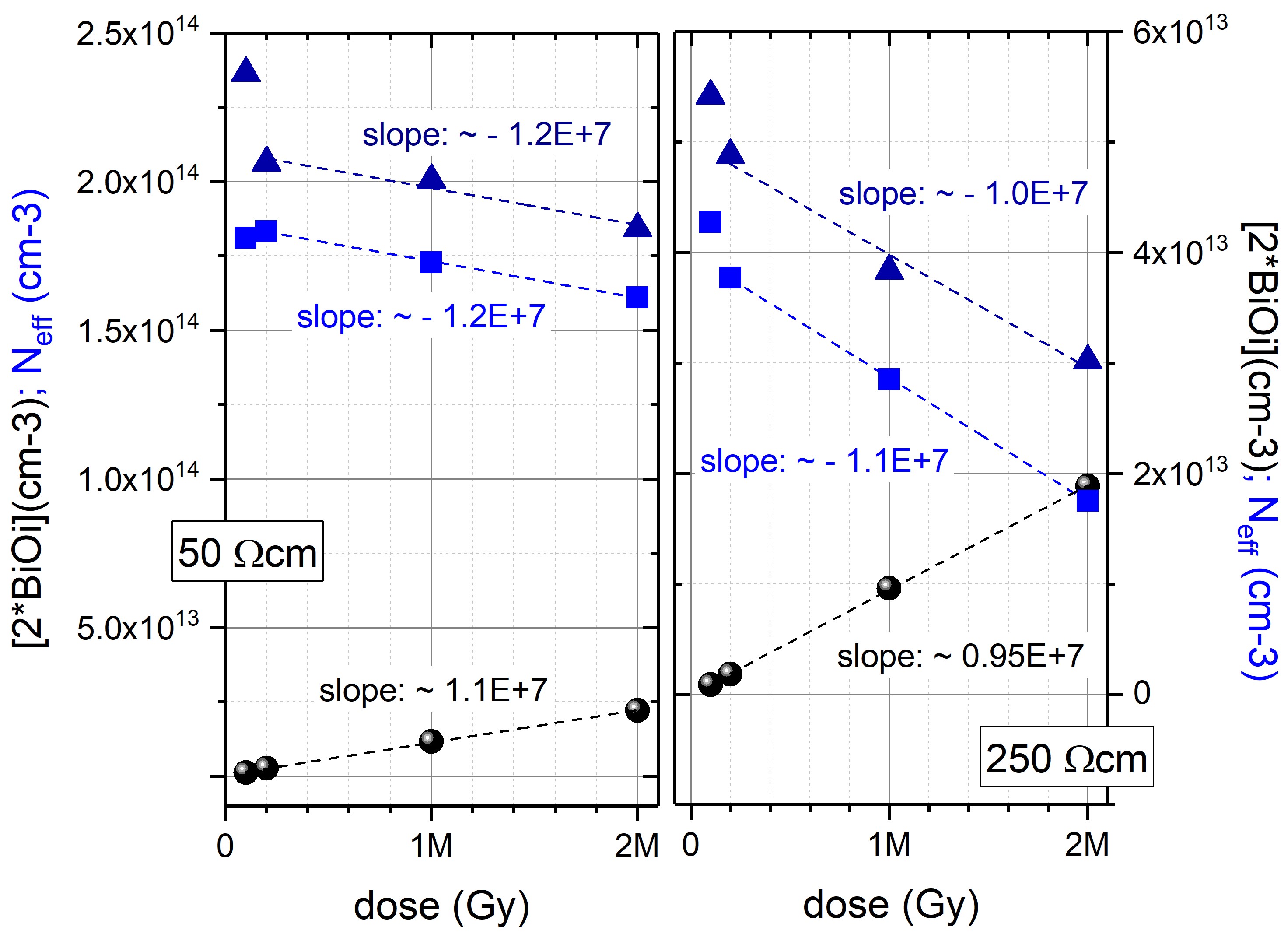}
    \caption{Evolution of twice the B$_\text{i}$O$_\text{i}$ concentration and the effective doping concentration \textit{N}$_\text{eff}$ with irradiation dose for $\gamma$-irradiated 50\,$\Omega$cm and 250\,$\Omega$cm EPI diodes. The values of the slope given in the plot are in units of Gy$^{-1}$cm$^{-3}$. Details are given in the text.} 
    \label{fig:concentr-Neff}
\end{figure}

Fig.\ref{fig:concentr-Neff} shows the correlation between the B$_\text{i}$O$_\text{i}$ concentration (taken twice) and the change in the effective doping concentration \textit{N}$_\text{eff}$ of the irradiated 50\,$\Omega$cm and 250\,$\Omega$cm diodes. For \textit{N}$_\text{eff}$ two different values are plotted in Fig. \ref{fig:concentr-Neff}. The values depicted by blue triangles were extracted from the \textit{C-V} measurements shown in Fig. \ref{fig:CV} that were performed at 253\,K and 1\,kHz. The \textit{N}$_\text{eff}$ values depicted by blue squares were extracted from \textit{C-V} measurements performed during the DLTS scan at 1\,MHz in the temperature range of B$_\text{i}$O$_\text{i}$  charge emission (T\,=\,108\,-\,130\,K). The slope of both curves is comparable and correlates very well with the change in B$_\text{i}$O$_\text{i}$ concentration for doses $\geq$\,200\,kGy. The observed correlation supports the assumption that the B$_\text{i}$O$_\text{i}$ formation is the main responsible mechanism for acceptor removal in irradiated boron-doped silicon. Thereby, the formation of one B$_\text{i}$O$_\text{i}$ would deactivate two active boron atoms. The presented results are in good agreements with similar measurements on proton irradiated epitaxial silicon diodes \cite{Liao2022IEEE}.   
%--------------------------------
%--------------------------------

\subsection{Analysis and modeling of TSC spectra}
\begin{figure}[b!]
    \centering
    \includegraphics[width=1\columnwidth]{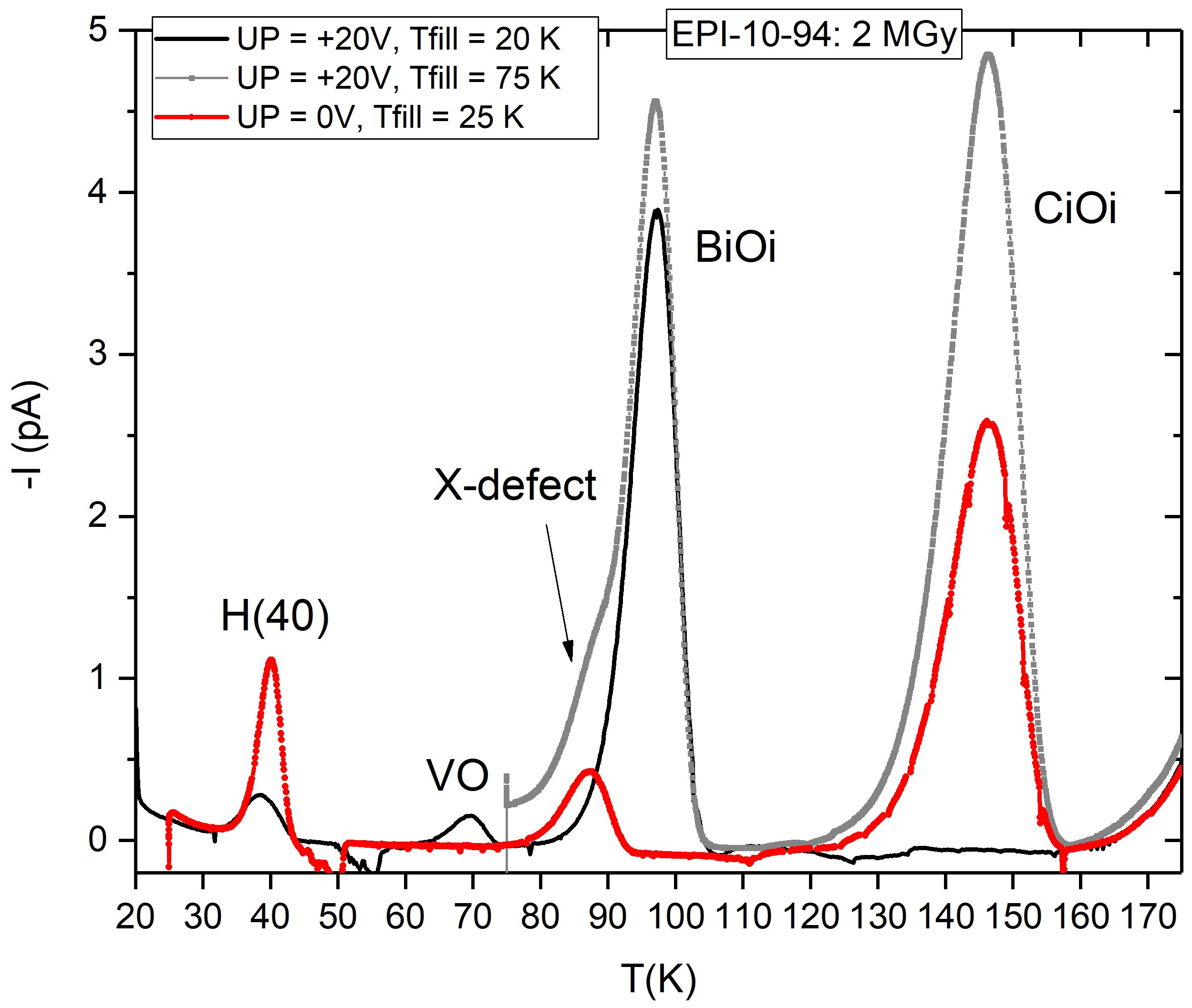}
    \caption{TSC spectra measured on a 2\,MGy $\gamma$-irradiated EPI diode with 250\,$\Omega$cm resistivity. The spectra differ in the applied pulse voltage UP as well as the temperature at which the defect filling took place \textit{T}$_\text{fill}$. The reverse bias was set to UR\,=\,-100\,V to fully deplete the diode. } 
    \label{fig:TSC}
\end{figure}

Besides DLTS investigations also TSC measurements in the temperature range of 20\,K to 220\,K were performed on the $\gamma$-irradiated EPI diodes. Fig. \ref{fig:TSC} shows as example the TSC spectra measured on the 250\,$\Omega$cm EPI diode $\gamma$-irradiated with 2\,MGy. Five prominent TSC peaks are revealed by these experiments, three belonging to hole traps labeled as  H(40), X-defect and C$_\text{i}$O$_\text{i}$ and two to electron traps, VO and B$_\text{i}$O$_\text{i}$.
The filling pulse voltage UP of the first two spectra in Fig.\, \ref{fig:TSC} was set to a forward bias of +\,20\,V while the filling temperature \textit{T}$_\text{fill}$ was varied. This was done since the electrical filling of defects like C$_\text{i}$O$_\text{i}$ and X-defect is temperature dependent when injecting with forward bias \cite{Pintilie2008APL, Liao2022IEEE}. Their TSC signal intensity starts to increase with increasing \textit{T}$_\text{fill}$. In Fig. \ref{fig:TSC} (grey line-square spectrum) the X-defect as well as the C$_\text{i}$O$_\text{i}$ are shown after carrier injection at 75\,K. The X-defect is partly overlapping with the B$_\text{i}$O$_\text{i}$ peak. 
%--------------------------------
%--------------------------------
\begin{figure}[b!]
    \centering
    \includegraphics[width=1\columnwidth]{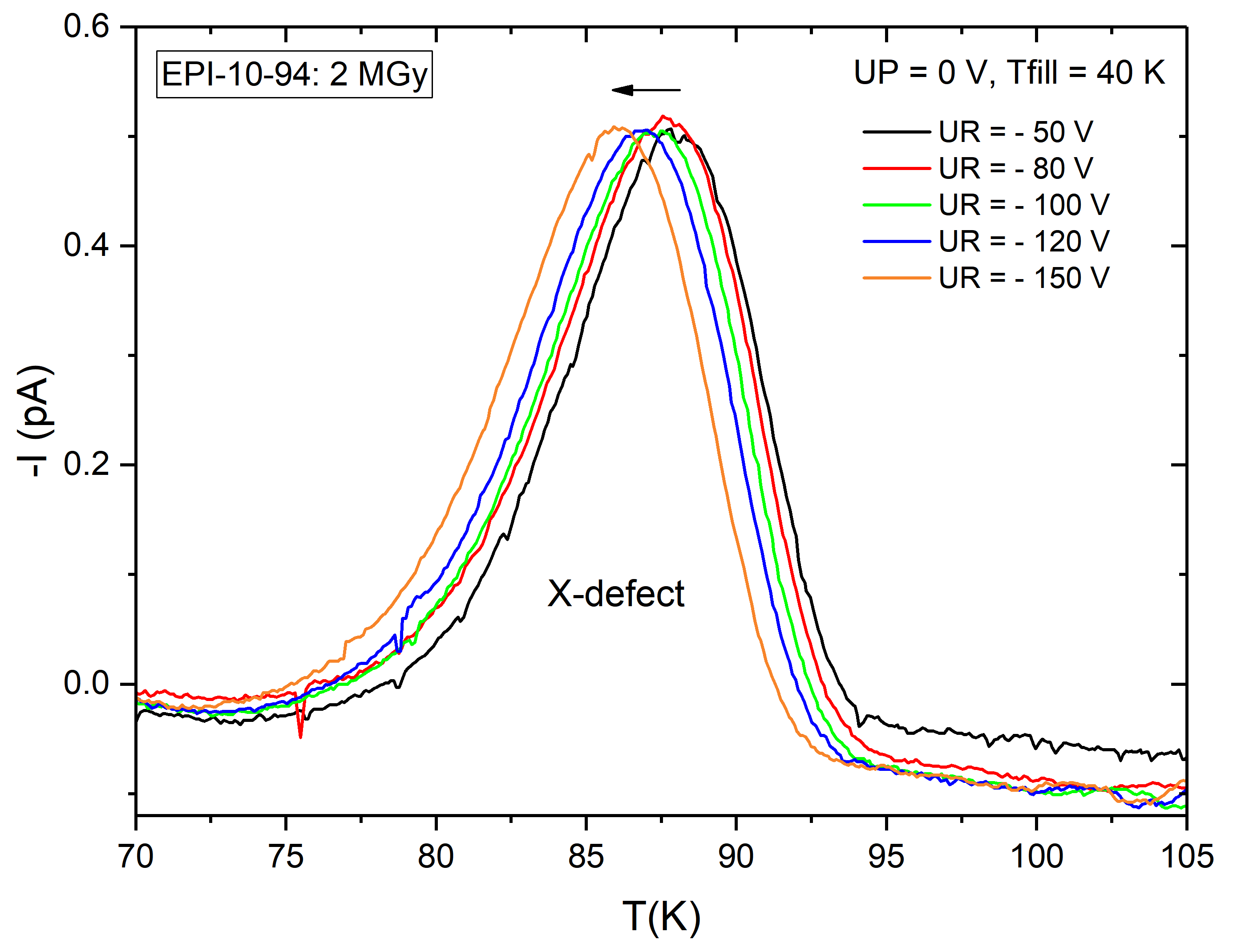}
    \caption{X-defect measured by TSC on a 2\,MGy $\gamma$-irradiated EPI diode with 250\,$\Omega$cm resistivity. As reverse bias during charge emission values from -50\,V to -150\,V were set.  } 
    \label{fig:TSC-X-def-bias}
\end{figure}
%--------------------------------
%\begin{figure}[b!]
%    \centering
%    \includegraphics[width=1\columnwidth]{TSC-250Ohm.png}
%    \caption{(a) Modeled TSC spectra using \textit{pytsc} of EPI diodes with 250\,$\Omega$cm resistivity $\gamma$-irradiated in the range of 1\,kGy to 2\,MGy. The defect parameters for the modeling were taken from the corresponding DLTS measurements of the four diodes. A reverse bias of -100\,V was set. (b) Experimentally measured TSC spectra of the same four diodes. The measurement conditions were: UR\,=\,-100\,V, UP\,=\,+\,20\,V, \textit{T}$_\text{fill}$\,$>$\,40\,K.} 
%    \label{fig:TSC-pytsc}
%\end{figure}
%--------------------------------
%--------------------------------
\begin{figure}[b!]
    \centering
    \includegraphics[width=1\columnwidth]{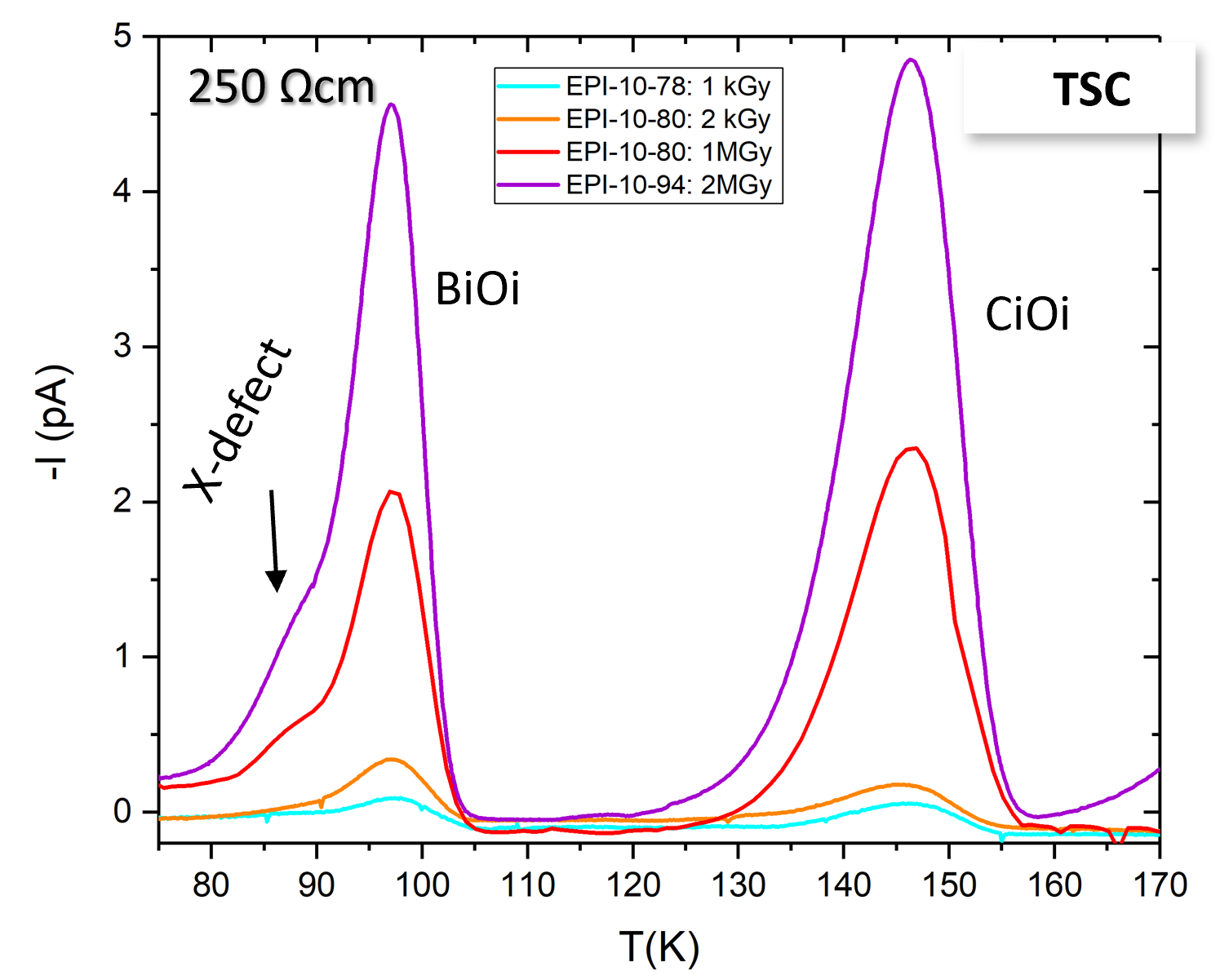}
    \caption{Measured TSC spectra of four EPI diodes with 250\,$\Omega$cm resistivity $\gamma$-irradiated in the range of 1\,kGy to 2\,MGy. The measurement conditions were: UR\,=\,-100\,V, UP\,=\,+\,20\,V, \textit{T}$_\text{fill}$\,$>$\,40\,K.} 
    \label{fig:TSC-plot}
\end{figure}
%--------------------------------
\begin{figure}[htb]
    \centering
    \includegraphics[width=1\columnwidth]{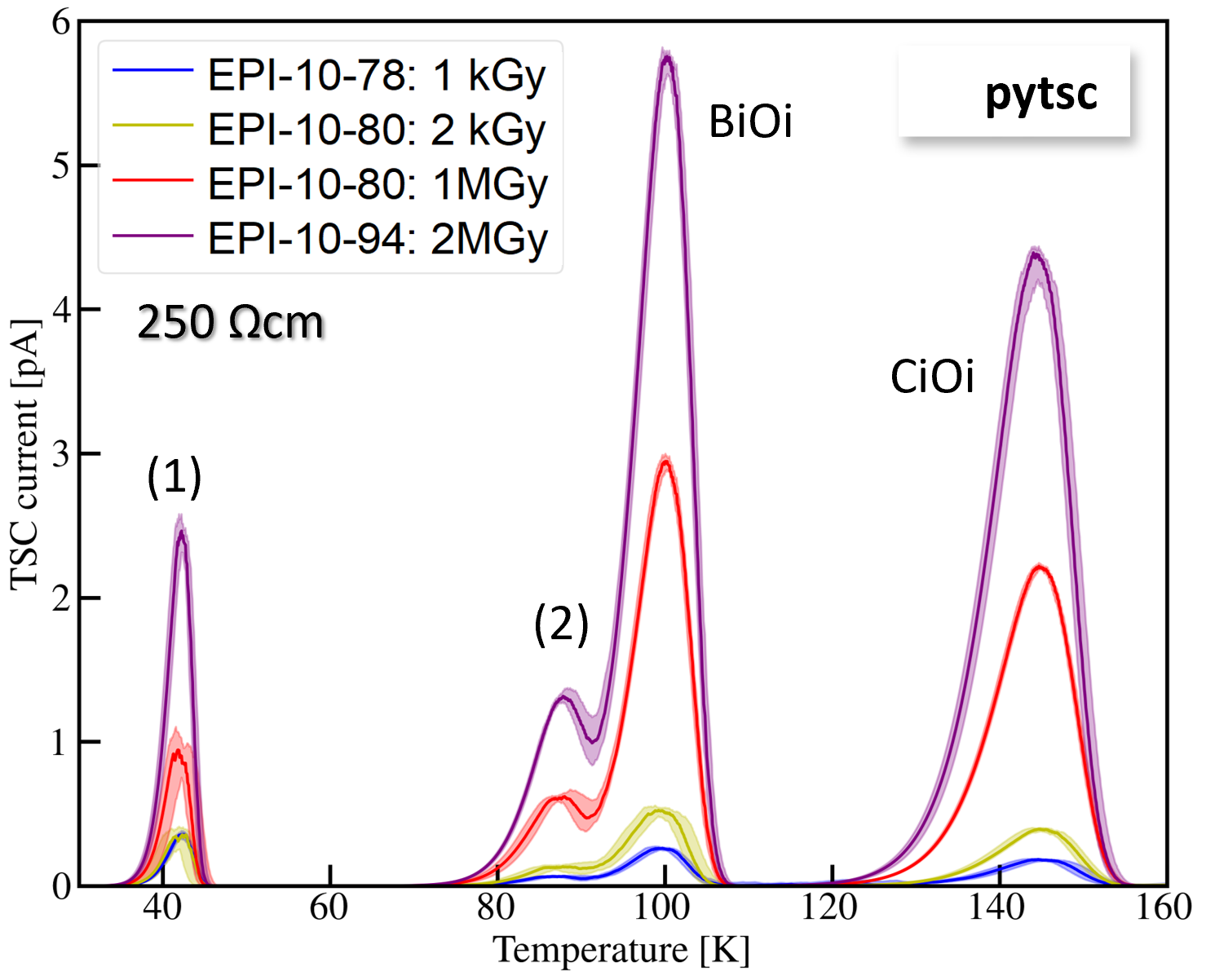}
    \caption{Modeled TSC spectra using \textit{pytsc} of EPI diodes with 250\,$\Omega$cm resistivity $\gamma$-irradiated in the range of 1\,kGy to 2\,MGy. The defect parameters for the modeling were taken from the corresponding DLTS measurements (illustrated in Fig.\,\ref{fig:250Ohm-DLTS}) of the four diodes. A reverse bias of -100\,V was set. Further details about the modulation are given in the text.} 
    \label{fig:pytsc-plot}
\end{figure}
%--------------------------------

In order to separate these peaks and distinguish the defect levels between electron and hole traps, TSC spectra with filling pulse UP\,=\,0\,V were also recorded. In this case only majority carriers were injected and the detected peaks H(40), X-defect and C$_\text{i}$O$_\text{i}$ can be assigned to hole traps. The B$_\text{i}$O$_\text{i}$ and the VO which are known to be electron traps \cite{Mooney1977, Pintilie2003APL} are not detected in this case (see red line-dotted spectra in Fig.\ref{fig:TSC}).\\
For the 50\,$\Omega$cm EPI diodes comparable measurements were performed (not shown here, see Ref.\cite{HimmerlichRD50}) and allowed also the clear identification of three peaks: the X-defect, the B$_\text{i}$O$_\text{i}$ and the C$_\text{i}$O$_\text{i}$ defect. Since in the lower temperature range the measured spectra of the 50\,$\Omega$cm diodes were dominated by a high setup-induced noise signal no information about defects with peak positions $<$ 60\,K can be given at this point. 

The clear assignment of the X-defect to a defect structure is still missing, although some specifications are already given in the literature \cite{Liao2022IEEE, Chuan2023electron}. Shortly summarized, it is known, that the X-defect is a hole trap with a strongly temperature dependent capture cross section, as also presented in the measurements discussed before. Furthermore, detailed studies within the RD50 collaboration on proton, neutron and electron irradiated \textit{p}-type silicon diodes stated that the X-defect can always clearly be identified by TSC for lower irradiation fluences ($<$\,\num{7e13}\,\unit{n_{eq}\per\centi\square\meter}), while for higher fluences it is usually not seen. Additionally, the peak position of the X-defect in the TSC spectra shows a strong field dependence that was usually explained by the Poole-Frenkel effect \cite{Liao2022IEEE, Chuan2023electron, Frenkel38}. 
This was also observed for the irradiated diodes presented in this paper. Corresponding TSC measurements are plotted in Fig.\ref{fig:TSC-X-def-bias}.  
Here the charge emission from the X-defect was measured after cooling down the sample under reverse bias of -100\,V, choosing a filling pulse of UP\,=\,0\,V and setting different reverse bias voltages UR when ramping up the temperature. With increasing the applied reverse bias UR in the range from -\,50\,V to -\,150\,V the X-defect peak position shifts to lower temperatures, indicating an electrical field dependence.

In order to promote the assignment of the defect levels detected in TSC, especially the X-defect, the measured TSC spectra were compared to modeled \textit{pytsc}-spectra as described in section \ref{sec:MatMet}. As defect parameters for the \textit{pytsc} modeling, data were taken that resulted from the DLTS experiments performed on the same sample. In Fig. \ref{fig:pytsc-plot} \textit{pytsc} spectra of the $\gamma$-irradiated 250\,$\Omega$cm EPI diode are presented. They can be compared to the measured TSC spectra illustrated in Fig.\,\ref{fig:TSC-plot}. The spectra of the 50\,$\Omega$cm EPI diodes (see Ref.\,\cite{HimmerlichRD50}) are comparable in the interpretation of the results to the 250\,$\Omega$cm ones and therefore not additionally illustrated in this publication. 
The 4 peaks in Fig. \ref{fig:pytsc-plot} correspond to peak (1), peak (2), as well as B$_\text{i}$O$_\text{i}$ and C$_\text{i}$O$_\text{i}$ identified in DLTS (see Fig. \ref{fig:50Ohm-DLTS} and \ref{fig:250Ohm-DLTS}). The peak amplitudes increase with increasing radiation dose. The same is observed in the TSC measurements (see Fig. \ref{fig:TSC-plot}), although there is a difference in the absolute peak heights. The modeled B$_\text{i}$O$_\text{i}$ and C$_\text{i}$O$_\text{i}$ peak positions are in very good agreement to the measured ones. The DLTS-Peak (2) gives in \textit{pytsc} a signature next to the B$_\text{i}$O$_\text{i}$ that strongly resembles the shoulder induced by the X-defect in TSC, while the position of DLTS-peak (1) correlates with the H(40) defect (see Fig. \ref{fig:TSC} and Fig. \ref{fig:pytsc-plot}). As mentioned in section \ref{subsec: DLTS} the DLTS defect characteristics of peak (1) are comparable to those of the I$_{2}$O defect, that points towards a correlation between the TSC H(40) peak and the I$_{2}$O defect. For peak (2) the comparison of our DLTS results with literature, as discussed before in section \ref{subsec: DLTS}, points to the donor state of the divacancy V$_{2}$(0/+). This defect is neutral before trapping a hole and positively charged afterwards.%In TSC for \textit{p}-type Si a filling of the defect states with  holes as majority carriers is done before applying a reverse bias and measuring the TSC emission current.
%Under these measurement conditions, after the hole injection a V$_{2}$(0/+) defect would be considered to be positively charged, therefore a lowering of the potential barrier due the Coulomb potential would be expected and for the carrier emission process a field-dependence due to the Poole-Frenkel effect is expected \cite{Zangenberg2002}. Such a field dependence is observed for the X-defect when varying the reverse bias applied during the TSC measurements. In future studies now the indications for assigning the X-defect to a di-vacancy structure must be finally proofed by ongoing experimental studies using also light injection experiments and annealing studies.
Therefore, according to the definition of the Poole-Frenkel effect that occurs only for coulombic centers, the emission rate of the defect should not vary with the applied electric field. However, a large field dependence is reported in literature for the  V$_{2}$(0/+) and correlated with phonon assisted tunneling \cite{Zangenberg2002, Zangenberg2005AP}. Also the X-defect showed a field dependence when varying the reverse bias applied during TSC measurements. These are indications to assign the X-defect to a di-vacancy structure while still further work is needed to fully consolidate or decline this attribution.  
%================================
\section{Summary and Conclusion}
The present work is summarizing defect studies on a set of epitaxial grown $^{60}$Co $\gamma$-irradiated \textit{p}-type silicon diodes of different resistivity. In this context, it is demonstrated that the observed changes in the effective carrier concentration of the diodes correlate, as expected, with the formation of the boron-related B{$_\text{i}$}O{$_\text{i}$} defect.
We clearly demonstrate that change of the effective space charge is double of the change of the B{$_\text{i}$}O{$_\text{i}$} defect concentration. This is fully consistent with describing the acceptor removal in boron doped p-type silicon as a process consisting of the removal of the shallow acceptor boron B{$_\text{s}$} and the subsequent formation of a B{$_\text{i}$}O{$_\text{i}$} defect with donor character, i.e. a positive space charge contributor.
Additionally, within this work, a continuative defect analysis method is presented by combining results from different spectroscopic methods (DLTS and TSC) by implementing a Python-based modeling software (\textit{pytsc}). Thereby, the defect parameters (activation energy, charge carrier capture cross section and defect concentration) obtained by DLTS are used for modeling the corresponding TSC spectra and comparing them with the experimental ones. It gives indications that the so-far unspecified X-defect, giving rise to a TSC peak that partially overlaps with that of B{$_\text{i}$}O{$_\text{i}$}, might be related to divacancies. However, further studies are needed and planned to fully confirm this assignment option.
%================================
\section{Acknowledgement}
The work was partly performed in the framework of the CERN RD50 and DRD3 collaborations. I. Pintilie acknowledge the funding received through IFA-CERN-RO 08/2022 project. 
%\begin{figure}[htb]
%    \centering
%    \includegraphics[width=1\columnwidth]{IV.png}
%    \caption{I-V measurements of $^{60}$Co gamma irradiated EPI diodes of different resistivity (top: 50\,$\Omega$cm, bottom: 250\,$\Omega$cm). Measurements were performed at -\,20\,$^{\circ}$\,C, with guard ring connected.} 
%    \label{fig:IV}
%\end{figure}

\bibliographystyle{elsarticle-num} 
\bibliography{literature}{}


%\begin{thebibliography}{00}


%\bibitem{}

%\end{thebibliography}
\end{document}